\documentclass[twocolumn,prb,nofootinbib,showpacs]{revtex4-1}
\usepackage{graphicx}
\usepackage{hyperref}
\usepackage{amsfonts}
\usepackage{amsmath,amssymb}
\usepackage{bm}

\begin{document}

\title{Induced spin texture in semiconductor/topological insulator heterostructures}
\author{Jimmy A. Hutasoit}
\email{jimmy.hutasoit@mail.wvu.edu} 
\affiliation{Department of Physics, West Virginia University, Morgantown, West Virginia 26506}
\author{Tudor D. Stanescu}
\affiliation{Department of Physics, West Virginia University, Morgantown, West Virginia 26506}

\begin{abstract}
We show that a semiconductor thin film can acquire a non-trivial spin texture due to the proximity effect induced by a topological insulator. The effect stems from coupling  to the topological surface states and is present even when the insulator is doped. We propose a semiconductor/topological insulator heterostructure as a device that allows measuring interface properties and probing  surface states in uncompensated samples. We also find that the topological insulator surface modes can be significantly  broadened and shifted by the presence of metallic contacts. 
\end{abstract}

\pacs{73.20.-r, 74.45.+c} 

\maketitle

\section{Introduction}

The three-dimensional time-reversal invariant topological insulators (TIs)~\cite{Kane2005,Bernevig2006,Fu2007a, Fu2007b,Moore2007,Roy2009}, or the so-called class AII TIs\cite{Ryu:2010fk}, have attracted considerable interest in recent years following the observation  of characteristic TI surface states in a family of strongly spin--orbit interacting  Bi--based materials~\cite{Hsieh2008,Xia2009,Hsieh2009}. The surface states of these topological insulators are robust against perturbations that do not break time reversal symmetry (e.g., disorder or many body interactions) and form a helical metal characterized by quasiparticles with the spin direction locked by the momentum. This helical metal is key to a series of exotic properties predicted to occur in structures containing TIs, such as axion electrodynamics~\cite{Qi2008,Essin2009}, Majorana fermions~\cite{Fu2008} and topological exciton condensates~\cite{Seradjeh2009}. These exciting proposals have created a flurry of activity dedicated to understanding the properties of the helical metal. For a review, see for example Refs. \onlinecite{Qi:2010fk,Hasan:2010uq}.

The main challenge to observing topological properties in real materials stems  from the fact that as--grown Bi$_2$Te$_3$ and Bi$_2$Se$_3$ have a significant intrinsic carrier density in the bulk conduction band. This does not prevent the observation of  surface states within the gap between the valence and conduction bands using angle resolved photoemission spectroscopy (ARPES)~\cite{Hsieh2008,Xia2009,Hsieh2009} or scanning tunneling spectroscopy (STS)~\cite{Urazhdin2002,Roushan2009}. However, characterization using other techniques, including  transport measurements, is severely restricted due to the difficulty of separating  the surface and bulk contributions. 

Moreover, no exotic topological property can be present in a non--insulating system. To fully realize the potential of TI materials, they have to be integrated into heterostructures containing superconductors, magnetic materials or trivial band insulators. Hence, a critical task is to characterize the properties of the topological interface states. While their existence is guarantied by topology, key properties of such states (e.g.,  dispersion, characteristic length scales, spin texture, and mixing with other types of in--gap states) depend on the details of the interface, such as the transparency on the interface or the presence of defects and charged impurities. Being able to characterize the interface states in the presence of bulk carriers would present significant practical advantages. Finally, as transport measurements involve placing metallic contacts on the TI surface, a  natural question that needs to be addressed concerns the fate of the surface states in the presence of these contacts. 

In this paper,  we propose a semiconductor thin film -- topological insulator heterostructure as a tool for studying the properties of the helical metal and  the dependence of these properties on the parameters of the interface. The key idea behind this proposal is  that the states localized at the interface between a TI and a semiconductor couple to the semiconductor bands and induce certain specific properties (e.g., a spin texture) due to proximity effect. Consequently, probing the semiconductor states provides a direct characterization of the interface states. 

The present proposal addresses two critical questions: (i) How can one disentangle the surface and bulk effects in doped TI samples? (ii) What are the properties of the interface in TI heterostructures  and how can one control these properties? 

The potential advantages of using the semiconductor-TI heterostructure proposed here to address these questions stem from the fact that semiconductor--based heterostuctures are, in general, easier to grow, while the properties  of the semiconductor thin film, and implicitly of the interface, can be measured by a variety of optical and transport probes. In addition, perfectly insulating TI samples are not required, as the proximity effect involves only surface TI states and quasi-two--dimensional semiconductor states. Furthermore, the strong dependence of the TI--semiconductor coupling on the thickness of the semiconductor film provides an extremely useful knob for tuning the strength of the proximity effect. 

To prove that the electronic  properties of the semiconductor thin film are modified in a very specific way by the coupling to the TI, we derive a generic expression for the proximity effect at the interface between a TI and a metal or semiconductor
and compare its predictions with the results of microscopic tight-binding calculations. The remarkable agreement between the two techniques reflects the robustness of the effect that we are describing. In fact, as verified explicitly, even changes in the Hamiltonian itself do not modify qualitatively  our conclusions, as long as the gap between the valence and the conduction TI bands  remains finite. 

We emphasize that our conclusions are based on calculations using two different approaches: (i) a microscopic tight--binding model  for a three-dimensional TI -- semiconductor (or metal) structure in the slab geometry, and (ii) an effective two--dimensional description of the interface using the standard  Green's functions formalism (for an introduction to this formalism, see, for example, Ref. \onlinecite{mahan}). 

The paper is organized as follows. The tight--binding description is presented in Sec. \ref{sec:tightmodel} and includes a four-band model for the the Bi--based TI's, a one band model for the metal/semiconductor and a coupling term. The simple TI model can be generalized to include more bands  and thus provide quantitative predictions about the interface properties. The other crucial ingredient required for obtaining quantitative results is the coupling Hamiltonian, Eq, (\ref{HTI}). The optimal values of the tunneling matrix elements could be obtained by  comparison with future experimental measurements. These values are critical for determining the strength of the proximity effect and depend on the microscopic details of the interface. In principle, diagonalizing numerically the tight--binding Hamiltonian should be enough for supporting our claims. However, an effective low energy theory of the interface provides further physical insight into the proximity effect and the changes that it induces in the properties of the surface states. The effective interface theory for a TI in contact with a metal is presented in Sec. \ref{sec:greenmetal}. Within this framework, we study the fate of the surface states in the presence of metallic contacts. We find that the surface modes are shifted and broadened by the coupling to a continuum, but do not loose spectral weight. We address the case of an infinite metallic plate, as well as finite size contacts. In Sec. \ref{sec:induced} we derive the effective theory for a semiconductor thin film in contact with a TI. In contrast with the case discussed in Sec. \ref{sec:greenmetal}, we now integrate out the TI degrees of freedom and identify the effect of the surface states on the semiconductor spectrum. Again, the results are compared with the numerical solution of the tight--binding model for the TI--semiconductor heterostructure. Our conclusions and the proposal for the experimental realization of the TI-semiconductor thin film device are presented in Sec. \ref{sec:conc}.

\section{Tight-binding model of a topological insulator heterostructure} \label{sec:tightmodel}

The tight-binding Hamiltonian has the generic form 
\begin{equation}
H = H_{TI} + H_{\rm band} +V. \label{Ham}
\end{equation}
The first term represents a four band low--energy effective TI model,
\begin{equation}
H_{TI} = \sum_{\alpha, i, j}\left(\epsilon_0^{(\alpha)}\delta_{ij} + t_{ij}^{(\alpha)}\right) c_{i\alpha}^{\dagger} c_{j\alpha} + c_{i \alpha}^{\dagger}\left(i\lambda_{ij} {\bm \delta}\cdot\hat{\bm \sigma}\right) c_{j\bar{\alpha}}, \label{HTI}
\end{equation}
where $\alpha, \bar{\alpha}\in\{1, 2\}$, $\alpha\neq \bar{\alpha}$ are band indices, ${\bm \delta}={\bm r}_j-{\bm r_i}$ and $\hat{\bm \sigma}=(\hat{\sigma}_x, \hat{\sigma}_y, \hat{\sigma}_z)$ are Pauli matrices. 

The basis $\psi_{\alpha \tau}$ for this model contains even ($\alpha=1$) and odd  ($\alpha=2$) parity combinations of $p$ orbitals with a mix of up and down spins~\cite{Zhang2009}. Here, $\tau = \,\, \Uparrow (\Downarrow)$ represents a pseudo--spin degree of freedom. The corresponding creation operators are $c_{i \alpha}^\dagger = (c_{i \alpha \Uparrow}^\dagger, c_{i \alpha \Downarrow}^\dagger)$. 

The model is defined on a rhombohedral lattice with the lattice parameters of Bi$_2$Se$_3$. The hopping parameters are non-zero only for nearest neighbor in--plane  and out--of--plane  hoppings,  which are given by $(t_1^{(1)},  t_1^{(2)}, \lambda_1) = (1.43, -2.95, 0.29)$ eV and $(t_2^{(1)},  t_2^{(2)}, \lambda_2) = (0.03, -0.04, 0.12)$ eV, respectively. In the long wavelength limit, Eq. (\ref{HTI}) reduces to the four--band effective model of Zhang {\it et al.}~\cite{Zhang2009}. 

The second term of the Hamiltonian, $H_{\rm band}$,  describes the conduction band of a metal or  the valence band of a semiconductor and contains only nearest neighbor hoppings on a hexagonal lattice that ensures simple matching conditions at the interface. These bands are double  spin degenerate. 

Finally, the third term of the Hamiltonian describes the coupling between the TI and the semiconductor (metal),
\begin{equation}
V=\sum_{i,j}\sum_{\alpha, \tau, \sigma}\left[ \tilde{t}_{ij}^{(\alpha \tau,\sigma)} c_{i \alpha\tau}^\dagger a_{j\sigma} + {\rm c.c.}\right], \label{V}
\end{equation}
where the fermion operators $c_{i \alpha\tau}$ and  $a_{j\sigma}$ operate in the Hilbert space of the TI and semiconductor (metal), respectively, and $\tilde{t}_{ij}^{(\alpha \tau,\sigma)}$ characterizes the transparency of the interface. Experimentally, these parameters can be modified by depositing a thin insulating layer at the interface. 

Assuming for simplicity  that the coupling parameters are real,  the coupling is described by four independent quantities, $ \tilde{t}_{ij}^{(\alpha \Uparrow,\uparrow)} = \tilde{t}_{ij}^{(\alpha \Downarrow,\downarrow)} = \tilde{t}_{\alpha}$ and $ \tilde{t}_{ij}^{(\alpha \Uparrow,\downarrow)} = \tilde{t}_{ij}^{(\alpha \Downarrow,\uparrow)} = \tilde{t}_{\alpha}^\prime$, where $i$ and $j$ represent  nearest neighbor sites from interface boundary layers of the TI and semiconductor (metal), respectively. We note that the dominant contributions to the basis states $\psi_{\alpha \tau}$ come from $p_z$ orbitals with spin parallel to the pseudo--spin~\cite{Liu2010}. Hence, we expect $|\tilde{t}_{\alpha}| \gg  |\tilde{t}_{\alpha}^\prime|$. The total tight-binding Hamiltonian is then diagonalized numerically for a slab geometry.

\section{Topological insulator surface states in the presence of metallic contacts} \label{sec:greenmetal}

\subsection{Infinite metallic plate}

To obtain a deeper understanding of the physics at the interface, we  develop an effective two--dimensional description of the relevant low--energy degrees of freedom (i.e., the TI surface states and the semiconductor or metallic bands). Specifically, we study the change of surface states induced by the coupling to a thick metallic plate placed on the surface of the TI. In this case, the effective model is obtained by integrating out the metallic degrees of freedom and projecting into the subspace spanned by the surface states. 

\begin{figure}[tbp]
\begin{center}
\includegraphics[width=0.48\textwidth]{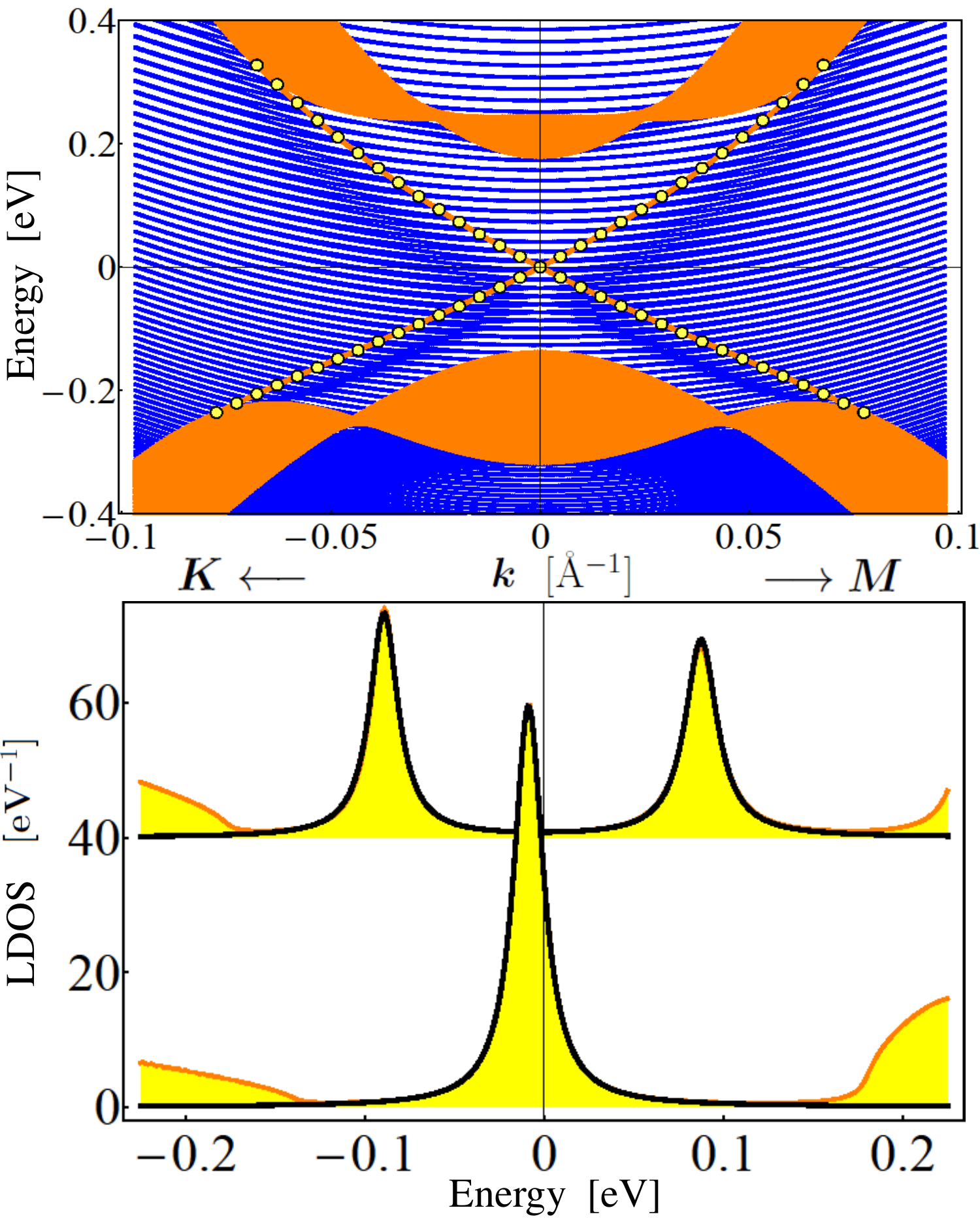}
\end{center}
\vspace{-5mm}
\caption{(Color online)  Top: Tight--binding spectrum of a TI slab with a planar metallic contact. Orange (light gray) represents uncoupled TI states, while blue (dark gray) corresponds to coupled TI--metal states. In the darker region within the bulk TI gap the spin degeneracy of the metallic states is lifted by coupling to surface TI states. The small circles (yellow) are calculated using the effective two-dimensional model (\ref{Hsurf}) for the surface states. Bottom: Local density of states at the TI--metallic contact interface for  ${\bm k} = (0,0)$ (lower curves) and ${\bm k} = (0.03 ,0)$ $\rm \AA^{-1}$ (upper curves; shifted for clarity).  Orange (light gray) filled lines represent tight--binding calculations, while black lines are solutions of the effective model (\ref{Ginv}).} \label{FIG1}
\vspace{-5mm}
\end{figure}

In the translation invariant case, where the length and width of the TI and metal are the same (taken to be infinity), we have
\begin{equation}
G^{-1}_{\lambda\lambda^\prime}({\bm k}, \omega) = \left[\omega  - \epsilon_\lambda({\bm k})\right]\delta_{\lambda\lambda^\prime} -\Sigma_{\lambda\lambda^\prime}({\bm k}, \omega), \label{Ginv}
\end{equation}
where $G^{-1}({\bm k}, \omega)$ is the inverse of the Green's function for the surface states and ${\bm k}$ the two-dimensional wave vector.  The energies $\epsilon_\lambda({\bm k})$, $\lambda=\pm$,  are the eigenvalues of the effective Hamiltonian for a free surface ($V=0$),
\begin{equation}
H_{\rm surf} = C k^2 \hat{\sigma}_0 + (A_0+A_2 k^2)[\hat{\bm \sigma}\times{\bm k}]\cdot \hat{z}, \label{Hsurf}
\end{equation}
where $\hat{\sigma}_0$ is the $2\times 2$ identity matrix and $\hat{z}$ is the unit vector perpendicular to the surface. We note that there is a one-to-one locking between the momentum and the spin, with no out-of-plane spin component. The parameters of the model were determined by the condition that $\epsilon_\pm({\bm k})$ match the tight--binding surface modes (see Fig. \ref{FIG1}): $C=13.33$ eV$\cdot$\AA$^2$,  $A_0=3.49$ eV$\cdot {\rm \AA}$, and $A_2=99.2$ eV$\cdot$\AA$^3$. 

{\it A priori}, we can expect that upon coupling to the metal, the spectrum of the surface states will be broadened. This is because even though the spectral profile of the surface states is trivial (i.e., is given by a Dirac $\delta$ function) viewed from the (quasi) two-dimensional point of view, the spectral profile of the metallic states is given by a broad continuous function due to its dependence on the momentum on the $z$ direction. Therefore, upon coupling, the surface states will inherit a non-trivial spectral density from the metallic states. Such phenomena have also been observed in particle physics, see, for example Refs. \onlinecite{Boyanovsky:2008bf,Boyanovsky:2009ke}.

In Eq. (\ref{Ginv}), the coupling to the metal is captured by the self-energy~\cite{Z2010a, Z2010b}
\begin{equation}
\Sigma_{\lambda\lambda^\prime}({\bm k}, \omega) = \sum_{\sigma, \nu} V_{\lambda, \nu\sigma}({\bm k})  G_\nu^{M}({\bm k}, \omega)  V_{\nu\sigma, \lambda^\prime}({\bm k}), \label{Sigma}
\end{equation}
where 
\begin{equation}
V_{\lambda, \nu\sigma}({\bm k}) = \langle \Psi_{\lambda}(\bm k)|V|\phi_{\nu\sigma}({\bm k})\rangle,
\end{equation}
are the matrix elements of the interaction Hamiltonian between TI surface states and metal states and 
\begin{equation}
G_\nu^{M}({\bm k}, \omega)=\left(\omega - E_\nu({\bm k}) +i\eta\right)^{-1},
\end{equation} 
is the Green's function for the metal. 

Explicitly carrying out the summation in Eq. (\ref{Sigma}), we have 
\begin{equation}
\Sigma_{\lambda\lambda^\prime}({\bm k}, \omega)= g_{\lambda\lambda^\prime}({\bm k}) ~\Gamma({\bm k}, \omega), \label{explicit}
\end{equation}
with 
\begin{equation}
\Gamma({\bm k}, \omega) = \frac{\omega-\xi_{\bm k}-\Lambda_{\bm k}}{\Lambda_{\bm k}} - i~\sqrt{1-\frac{(\omega-\xi_{\bm k}-\Lambda_{\bm k})^2}{\Lambda_{\bm k}^2}}. \label{Gamm}
\end{equation}
Here, $\xi_{\bm k}$ is the lowest energy of the metal at ${\bm k}$,  $\Lambda_{\bm k}$  is the half--bandwidth at the same wave vector and $\xi_{\bm k} \leq \omega \leq \xi_{\bm k} +2 \Lambda_{\bm k}$ (i.e., the energy is within the metallic band). 

To obtain the matrix $g_{\lambda\lambda^\prime}$, we note that in the vicinity of the $\Gamma$ point, ${\bm k}=(0,0)$, the four components of the surface states at the boundary $z=z_b$ take the form
\begin{equation}
\Psi_{\lambda}({\bm k}; z_{b}) = [u(k), iv(k), \lambda u(k) e^{i\varphi_{\hat{k}}},  -i \lambda v(k) e^{i\varphi_{\hat{k}}}]^T, \label{psi} 
\end{equation}  
with the real amplitudes $u(k)$ and $v(k)$ depending only on $|{\bm k}|$ and the phase $\varphi_{\hat{k}}$ determined by the direction $\hat{k}$ of the wave vector. If, for simplicity,  we neglect the dependence on $\tilde{t}_\alpha^\prime$, the coupling becomes diagonal  and we have
\begin{equation}
g_{\lambda\lambda^\prime}= \frac{4(u^2\tilde{t}_1^2 + v^2\tilde{t}_2^2)}{\Lambda_{\bm k}} \delta_{\lambda\lambda^\prime}.
\end{equation}

We conclude that the self--energy in Eq. (\ref{Ginv}) is  proportional to the weight of the surface states at the boundary and to the square of the coupling matrix elements. The real part of the self-energy $\Sigma$ shifts the surface modes, while the imaginary part broadens the spectrum.  This behavior is illustrated in Fig. \ref{FIG1}.  

In Fig. \ref{FIG1}, the upper panel shows the spectrum of a thick TI--metal slab with coupling $\tilde{t}_1=\tilde{t}_2=0.1$ eV and, for comparison, the spectrum of an uncoupled TI slab, both obtained using the tight-binding model. The energies of the surface states can also be obtained by diagonalizing the effective surface Hamiltonian (\ref{Hsurf}) (small circles in Fig. \ref{FIG1}). In the presence of TI-metal coupling, the surface states hybridize with metallic states and do not longer form sharply defined modes. In turn, the spin degeneracy of the metal states is lifted (more on this in the next section). The spectral weight distribution at the interface is shown in the lower panel of Fig. \ref{FIG1}. We note the remarkable agreement between the tight--binding (filled lines) and the effective model (black lines) calculations at energies within the bulk TI gap. The positions of the peaks are given by the real parts of the poles of the Green's function (\ref{Ginv}), while the widths are given by the imaginary parts.  The non--zero local density of states at the boundary generated by bulk TI states can be captured only by the tight--binding approach. We also note that the total weight of an in-gap resonance is independent of the coupling strength, as long as the peak is inside the gap. When the effective coupling $g_{\lambda\lambda}$ is of the order of the gap, the contribution from bulk TI states cannot be neglected and the effective theory is no longer valid.  

Before moving on to the case of finite size contacts, we would also like to note that the broadening of the surface modes by the metallic contact has also been observed in a different tight-binding model \cite{Zhao:2010fk}.

\subsection{Finite size contacts}

In transport measurements, the size of the contact is finite and thus, translation invariance is broken. Therefore, let us consider the case of a thin contact of length $L_y^{\rm metal} \ll L_y^{\rm TI}$, where $L_y^{\rm TI}$ is the length of the TI, taken to be finite. The dimensions in the $x$ direction are the same. Since translational invariance is broken, momentum in the $y$ direction is no longer conserved and the self-energy becomes a matrix in terms of the discrete momentum $k_y$ 
\begin{eqnarray}
\Sigma_{\lambda\lambda^\prime}(k_y, k_y', k_x, \omega) = \sum_k \,&& \frac{D\left(\tfrac{2 \,k \, \pi}{L_y^{\rm metal}} - k_y\right) D\left(\tfrac{2 \,k \, \pi}{L_y^{\rm metal}} - k_y'\right)}{L_y^{\rm metal}} \nonumber \\ 
&&\Sigma_{\lambda \lambda'}\left(\tfrac{2 \,k \, \pi}{L_y^{\rm metal}}, k_x, \omega\right),
\end{eqnarray}
where $\Sigma_{\lambda \lambda'}\left(\tfrac{2 \,k \, \pi}{L_y^{\rm metal}}, k_x, \omega\right)$ is given by Eq. (\ref{explicit}) and the ``diffraction" function is given by
\begin{equation}
D(k_y - k'_y) =  \frac{2\, \sin\left(L_y^{\rm metal}\,(k_y - k'_y) \slash 2\right)}{k_y - k'_y}.
\end{equation}

The result for the band shift in this case is identical to the translation invariant case, but with rescaled coupling constants,
\begin{equation}
\tilde{t}_{\rm eff} = \tilde{t} \, \sqrt{L_y^{\rm metal}/L_y^{\rm TI}}.
\end{equation}
The peak positions (see Fig. \ref{FIG1}) are still given by the real part of the poles of the Green's function (\ref{Ginv}), $\omega_{\pm}(k_x,  \langle k_y\rangle)$, with $ \langle k_y\rangle$ the average $k_y$ component. However, the broadening acquires an extra contribution  $\partial \omega_{\pm}/\partial  \langle k_y\rangle /L_y^{\rm TI}$  coming from the momentum uncertainty.

This result can be generalized to the case where the contact in the $x$ direction is also finite in a straight-forward manner. We conclude that in the case of finite contact, when the size of the metal is negligible compared to the size of the TI, the shift and the broadening of the spectrum of the surface states are negligible.

\section{Induced spin texture in semiconductor/topological insulator  heterostructures} \label{sec:induced}

Let us now consider a TI-semiconductor heterostructure consisting of a thin semiconductor film on top of a thick TI slab. A very thin insulating layer at the interface allows us to control the coupling between the two subsystems. In this work, we consider an ``ideal'' interface (i.e., we do not include effects due to defects, charged impurities, or lattice mismatch).  Our analysis  is intended to be a proof of concept in support of the idea that a TI-semiconductor thin film heterostructure can be used for (i) characterizing the TI surface states in the presence of bulk carriers, and (ii) studying the physics of the interface. 

The semiconductor film can be described by an effective theory analogous to Eq. (\ref{Ginv}), 
\begin{equation}
G_{\nu\sigma~\nu^\prime\sigma^\prime}^{-1}({\bm k}, \omega) = [\omega - E_{\nu}({\bm k})]\delta_{\nu\nu^\prime}\delta_{\sigma\sigma^\prime} - \Sigma_{\sigma\sigma^\prime}^{\nu\nu^\prime}({\bm k}, \omega), 
\end{equation}
with 
\begin{equation}
\Sigma_{\sigma\sigma^\prime}^{\nu\nu^\prime}({\bm k}, \omega) = \sum_{\lambda} V_{\nu\sigma,\lambda}({\bm k})\frac{1}{\omega - \epsilon_\lambda +i\eta} V_{\lambda,\nu^\prime\sigma^\prime}({\bm k}), 
\end{equation}
where $V_{\nu\sigma,\lambda}$ is the coupling matrix element between a semiconductor state with energy $E_{\nu}$ and spin $\sigma$ and a TI surface state with energy $\epsilon_{\lambda}$. In a thin film, the semiconductor bands split into subbands indexed by $\nu$ and separated by $\Delta E \propto 1/m^* L_z^2$, where $m^*$ is the effective mass and $L_z$ the film thickness. The semiconductor has to be chosen so that the top valence subband lies within the TI bulk gap. We assume that the uncoupled semiconductor bands are spin degenerate (i.e., the spin-orbit coupling in the semiconductor is negligible). However, when the coupling to the TI is turned on, $\Sigma_{\sigma\sigma^\prime}$ acquires off-diagonal contributions, or in other words, an effective spin-orbit coupling is induced by the proximity effect. 

To understand qualitatively this effect, we focus on the top valence subband $\nu=\nu_0$ and we neglect the inter--band coupling  
\begin{equation}
\Sigma_{\sigma\sigma^\prime}^{\nu\nu^\prime} \approx  \Sigma_{\sigma\sigma^\prime}^{(\nu)}\delta_{\nu\nu^\prime}.
\end{equation} 
This becomes exact in the limit $m^*\rightarrow 0$, $L_z\rightarrow 0$, but our final conclusions are independent of this approximation.  Taking into account the form of the surface states at the interface, Eq. (\ref{psi}), the relevant coupling matrix elements are
\begin{equation}
V_{\lambda,\nu_0\uparrow} = \phi_{\nu_0}\left[u(\tilde{t}_1 + \lambda \tilde{t}_1^\prime e^{i\varphi}) + i v (\tilde{t}_2 -\lambda\tilde{t}_2^\prime  e^{i\varphi})  \right]  \label{Vup},
\end{equation}  
and $V_{\lambda,\nu_0\downarrow} =  e^{i\varphi}V_{\lambda,\nu_0\uparrow}^*$, where $\phi_{\nu_0}$ is the wave function of the semiconductor state at the interface. 

The induced spin texture is determined by the structure of the matrix $\Sigma_{\sigma\sigma^\prime}^{(\nu_0)}$. 
\begin{figure}[tbp]
\begin{center}
\includegraphics[width=0.48\textwidth]{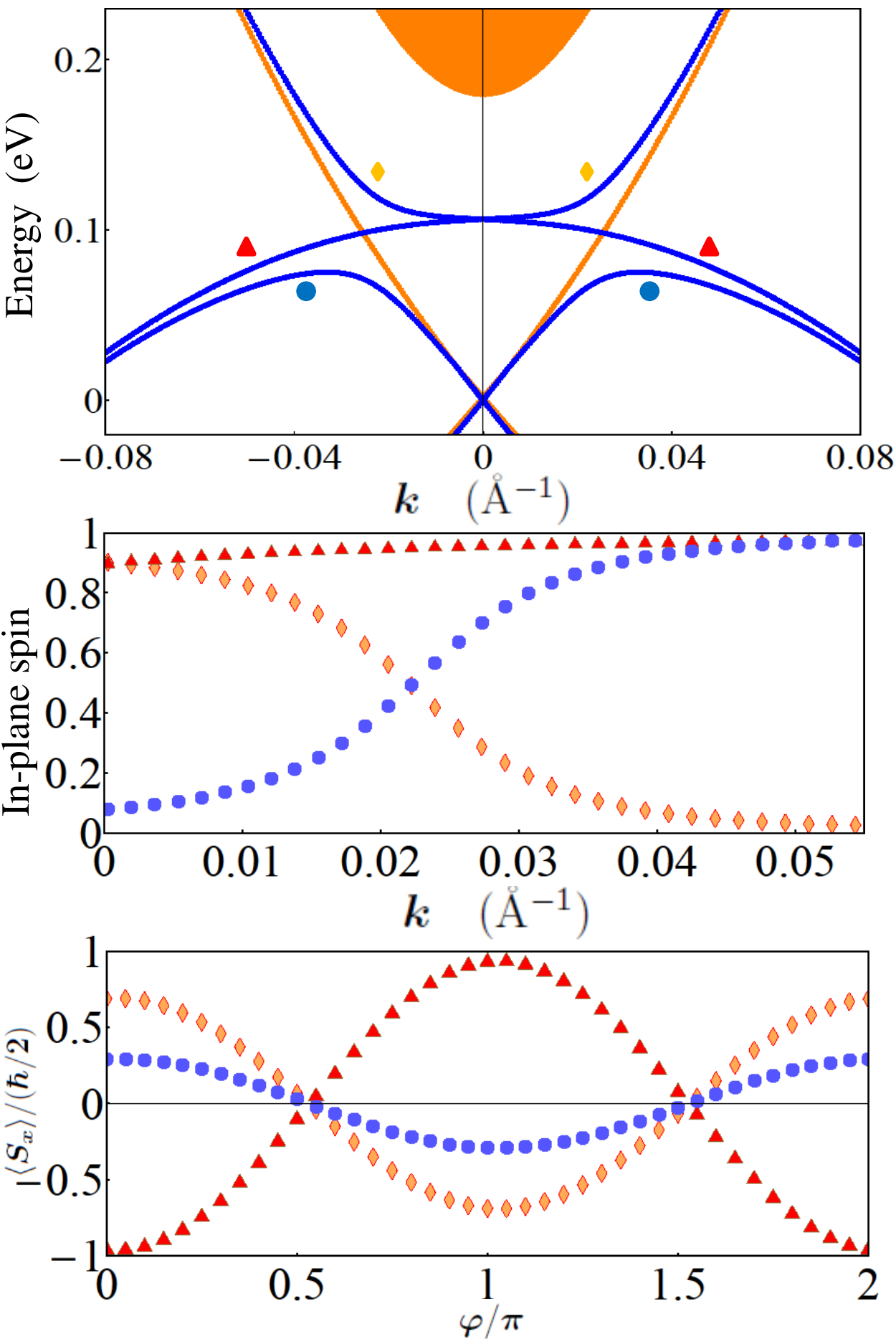}
\vspace{-5mm}
\end{center}
\caption{(Color online) Top: Spectrum of a TI-semiconductor thin film heterostructure.  Orange (light gray) represents TI bulk states and surface states localized near the free boundary. The blue (dark gray)  lines are coupled modes between TI interface states and top valence band states. Middle: Induced in-plane spin inside the semiconductor corresponding to the three coupled modes from the top panel.  The  out-of-plane components of the induced spin are negligible (less than $10^{-3}$).  Bottom: Dependence of the $x$ component of the induced spin on the direction in $k$ space.} \label{FIG2}
\vspace{-5mm}
\end{figure}
Taking into account the properties of the matrix elements $V_{\lambda,\nu_0\sigma}$ we have 
\begin{equation}
\Sigma_{\uparrow\uparrow}^{(\nu_0)} = \Sigma_{\downarrow\downarrow}^{(\nu_0)}, \qquad {\rm and} \qquad \Sigma_{\uparrow\downarrow}^{(\nu_0)} = \left[\Sigma_{\downarrow\uparrow}^{(\nu_0)}\right]^*. \label{Sigupdown}
\end{equation} 
As a result, the semiconductor states acquire a spin structure characterized by non-vanishing in-plane spin and zero out-of-plane component. This is a direct consequence of the helical pseudo--spin structure of the TI surface states. In particular, one can also verify that had the surface states had an out-of-plane component, the induced spin would have had an out-of-plane component as well. Therefore, for models with warped Dirac cone, in which away from the $\Gamma$ point  Eq. (\ref{psi}) is no longer valid, out-of-plane pseudo-spin and induced spin components are generated. 

The spin structure of the semiconductor thin film is illustrated in Fig. \ref{FIG2} using a tight-binding calculation for a heterostructure characterized by $\tilde{t}_1=0.1$ eV, $\tilde{t}_2=0.15$ eV, and $\tilde{t}_{\alpha}^\prime=0$. The spin degenerate valence band hybridizes with the TI surface states resulting 
three coupled modes (upper panel). For a given mode, the value of the in-plane spin (middle panel) is practically equal to the total spectral weight inside the semiconductor. The orientation of the in--plane spin is determined by the momentum direction (bottom panel). 

Near the $\Gamma$ point, the spin structure of the top modes can be described by an effective Rashba-like spin orbit coupling. The effective Rashba coefficient can be read from the effective model and for $\tilde{t}_1=\tilde{t}_2 = t$, it is simplified to
\begin{eqnarray}
A \approx  \frac{A_0 + A_2 \, k_{\parallel}^2}{2} \, \left(1 - \frac{\tfrac{2 \pi^2}{L_z m^*} - \mu}{\sqrt{(\tfrac{2 \pi^2}{L_z m^*} - \mu)^2 + \tfrac{4t^2}{\Lambda} \tfrac{2 \pi^2}{L_z m^*}}}\right),
\end{eqnarray}
where $\mu$ is the chemical potential and $\Lambda$ is given by twice of the lattice hopping parameter in the semiconductor.

We note that $\tilde{t}_{\alpha}^\prime \neq 0$ (i.e., having each pseudo-spin orientation coupled to both spin-up and spin-down) does not induce out-of-plane spin polarization, but rather generates an anisotropic structure characterized by the vanishing of the effective spin-orbit coupling away from the $\Gamma$ point.

\section{Conclusions} \label{sec:conc}

In this article, we have studied the proximity effects of the interface between TI and metal or semiconductor. The proximity effects cause a shift and broadening of the spectrum of the TI surface states, while lifting spin degeneracy on the metal or semiconductor side. In particular, by probing the induced spin texture in the semiconductor, one can learn about the properties of the TI surface states in more details, even in the excessive presence of the TI bulk carrier. Therefore, we propose a semiconductor thin film/TI heterostructure as a device for studying the properties of the TI surface states.

The induced spin texture can be probed experimentally using, for example,  spin-resolved ARPES or STS. Identifying the dispersion of the coupled modes provides direct information about the coupling at the interface. In particular, the strength of the coupling matrix elements can be determined as a function of the thickness of the interface insulating layer. The induced spin texture can also be probed using optical measurements. For optical measurements on Bi$_2$Se$_3$-semiconductor heterostructures, it is convenient to have the conduction band about 1 eV above the valence band, within the energy window characterized by a gap in the TI spectrum at low wave vectors~\cite{Xia2009,Zhang2009}. This will prevent the coupling between the conduction band and the TI, which can generate broadening of the energy levels. 

This proposal should complement direct measurement of the spin properties of the surface states such as that found in Ref. \onlinecite{Souma:2011uq} and should play a role in resolving the puzzle concerning the spin structure of TIs (see, for example, Ref. \onlinecite{Yazyev:2010vn} vs. Ref. \onlinecite{Zhang:2010kx}).

In the present study, we have considered  the case of an ideal interface. Future work is required for addressing problems such as the presence of interface defects and impurities, or the effects of the  lattice mismatch. These studies of the TI-semiconductor heterostructure will also help understanding key properties of TI-superconductor and TI-ferromagnet interfaces  that play a critical role in realizing many of the exotic properties of topological insulators.

\begin{acknowledgements}
We would like to thank our colleagues at WVU, in particular Sergei Urazhdin and  Alan Bristow, for their insightful comments. This work is supported in part by West Virginia University start-up funds.
\end{acknowledgements}

\vspace{-5mm}

\bibliography{References,contactsREF}
\end{document}